\documentclass[journal=jctc,manuscript=article]{achemso}
\usepackage[utf8]{inputenc}

\usepackage{booktabs}
\usepackage[version=3]{mhchem} 
\usepackage{xcolor}
\usepackage{comment}
\usepackage{multirow}
\usepackage{tablefootnote}
\usepackage{ulem}
\usepackage{graphicx}
\usepackage{array}
\usepackage{caption}
\usepackage{subcaption}
\usepackage{tikz}
\definecolor{brightpink}{RGB}{231, 84, 128}
\definecolor{brightpink}{RGB}{255, 105, 180}
\usepackage{hyperref}

\usepackage{xr}
\makeatletter

\SectionNumbersOn

\author{Kham Lek Chaton}

\author{Markus Meuwly} \altaffiliation{Department of Chemistry,
    Brown University, USA} \affiliation[University of
    Basel]{Department of Chemistry, University of Basel,
    Klingelbergstrasse 80, CH-4056 Basel, Switzerland.}
  \email{m.meuwly@unibas.ch}

\title{Machine Learning-Based Enhancements of Empirical Energy
  Functions: Structure, Dynamics and Spectroscopy of Modified
  Benzenes}


\begin{document}
\date{\today}

\begin{abstract}
The effect of replacing individual contributions to an empirical
energy function are assessed for halogenated benzenes (X-Bz, X = H, F,
Cl, Br) and chlorinated phenols (Cl-PhOH). Introducing electrostatic
models based on distributed charges (MDCM) instead of usual
atom-centered point charges yields overestimated hydration free
energies unless the van der Waals parameters are
reparametrized. Scaling van der Waals ranges by 10 \% to 20 \% for
three Cl-PhOH and most X-Bz yield results within experimental error
bars, which is encouraging, whereas for benzene (H-Bz) point
charge-based models are sufficient. Replacing the bonded terms by a
neural network-trained energy function with either fluctuating charges
or MDCM electrostatics also yields qualitatively correct hydration
free energies which still require adaptation of the van der Waals
parameters. The infrared spectroscopy of Cl-PhOH is rather well
predicted by all models although the ML-based energy function performs
somewhat better in the region of the framework modes. It is concluded
that refinements of empirical energy functions for targeted
applications is a meaningful way towards more quantitative
simulations.
\end{abstract}

\section{Introduction}
The success of empirical energy functions rests on two ingredients: i)
their functional dependence is sufficiently simple to allow rapid
evaluation for stable and long-time ($\mu$s or longer) simulations of
large systems ($10^6$ atoms or larger) and ii) their acceptable
accuracy in modelling intra- and intermolecular interactions which
allows qualitative or semi-quantitative molecular-level studies,
depending on the observable(s) considered and the agreement with
experiments sought. While existing force fields exploit new hardware
to simulate ever larger systems and longer timescales,\cite{shaw:2010}
increased computational power also provides an opportunity to include
more detail to refine force fields that are still applicable to
systems and timescales of chemical or biochemical interest. Crucial to
this task is determining to what extent increasing the detail of each
term provides greater accuracy in describing the intermolecular
interactions.\cite{MM.ffs:2024} At the same time, the final model
needs to be widely applicable with the possibility to parametrize it
even for highly heterogeneous chemical systems, such as ionic
liquids.\cite{MM.eutectic:2022,MM.eutectic:2024}\\

\noindent
Empirical energy functions have a long and successful history as
exemplified by the widely used parametrizations are the
CHARMM,\cite{charmmFF} Amber,\cite{amber} Gromos,\cite{gromos} and
OPLS\cite{opls} energy functions. Such energy functions were and are
being successfully applied to study proteins,\cite{huang:2014} nucleic
acids,\cite{nilsson:1986} organic crystals,\cite{price:2014} or
materials,\cite{deringer:2020} to name only a few applications. One of
the broadly applicable empirical energy functions is the CGenFF
parametrization which is the CHARMM General Force
Field.\cite{cgenff:2020,cgenff:2024} This energy function is available
as an additive and a polarizable variant using the Drude-oscillator
approach.\\

\noindent
On the other hand, most terms in an empirical energy function are
first-order approximations to more physically meaningful terms. One
example concerns harmonic, quadratic terms for chemical bonds which
are suitable to describe equilibrium structures of molecules and
small-amplitude vibrations within a normal mode
approximation. However, for anharmonic vibrations one needs to go
beyond such an approximation and include higher-order terms (cubic,
quartic) such as done, e.g., in the Merck Molecular Force Field
(MMFF).\cite{halgren:1996} There are alternative possibilities such as
replacing the harmonic terms by Morse-oscillators or machine-learned
energy functions using reproducing kernels or neural
networks.\cite{MM.review:2020,MM.review:2022,MM.rkhs:2017,MM.review:2023}
For studying the infrared spectroscopy of small molecules in
heterogeneous environments and comparing with experiments to provide
interpretations and understanding, the simplest parametrizations are
clearly insufficient.\cite{MM.mbco:2003,MM.mbco:2004} Similarly, to
characterized energy flow in molecules, the coupling between different
internal degrees of freedom needs to be captured considerably more
precisely for meaningful analysis and potentially predictive
simulations.\cite{MM.cn:2011,MM.nma:2014} In all these cases,
replacing specific terms in the empirical energy functions by more
appropriate functional forms will enhance the performance and improve
the quality of the simulations carried out with these energy
functions. \\

\noindent
Along similar lines, atom-centered point charges as an approximation
to the more general multipolar expansion of electrostatic interactions
can be replaced by, for example, extended charge
models,\cite{jorgensen:2012,Stone:1981,sokalski:1983,piquemal:2007,shaik:2008}
atom-centered
multipoles,\cite{MM.mtp:2012,MM.mtp:2013,price:2006,piquemal:2006,joubert:2002}
or off-centered distributed point
charges.\cite{MM.dcm:2014,MM.mdcm:2017,MM.mdcm:2020,MM.fmdcm:2022,MM.kmdcm:2024,gao:2014}
Such more elaborate models for the electrostatics capture the
anisotropy of the charge distribution around a particular interaction
site. This is, for example, relevant when representing lone pairs as
for water or $\sigma-$holes for halogenated
compounds.\cite{clark:2007,MM.mtp:2016} It is also possible to yet
include changes in the electrostatics as the structure of a molecule
changes along a MD trajectory. Such ``fluctuating charges'' or
``fluctuating multipoles'' are also capable of encapsulating bond
polarization\cite{price:1990,Reynolds:1992,stone:1995,MM.mbco:2008,Kim:2019,MM.fmdcm:2022,MM.kmdcm:2024}
or qualitatively describing charge reorganization for bond breaking
and bond-formation.\cite{MM.co2:2024}\\

\noindent
The present work explores the effects of replacing bonded and
electrostatic interactions in the CGenFF empirical energy functions
for halogenated benzenes and Chloro-phenol with the chloride atom at
its $o-$, $p-$, and $m-$positions relative to the hydroxyl. First, the
computational methods are presented, followed by structural,
spectroscopic and thermodynamic properties.\\

\section{Methods}
\subsection{Atomistic Simulations}
All simulations were performed using
CHARMM\cite{Charmm-Brooks-2009,MM.charmm:2024}. The off-entered
minimal distributed charge model MDCM were handled through the
DCM\cite{MM.dcm:2014,MM.mdcm:2017} module in CHARMM, and the
PhysNet-trained PES\cite{MM.physnet:2019} was run through
pyCHARMM.\cite{MM.physnet:2023,pycharmm:2023}\\

\noindent
Simulations for all solutes were carried out in a $32^3$ \AA\/$^3$
cubic water box using the TIP3P\cite{jorgensen:1983} water
model. There were a total of 1695 and 1696 atoms for the Halobenzene
(X-Bz) and fluoro-substituted phenol (F-PhOH) systems,
respectively. The systems were initially minimized using 500 steps of
Steepest Descent (SD) followed by 1000 steps of Adopted Basis
Newton-Raphson (ABNR) minimization. Subsequently, a heating and
equilibration step in the $NVT$ ensemble for 125 ps with timestep of
$\Delta t = 1$ fs was employed. The Nose-Hoover thermostat was used to
maintain the temperature around 303.15 K. This was followed by 1 ns
production simulations in the $NPT$ ensemble using the Leapfrog Verlet
integrator with a timestep of $\Delta t = 0.2$ fs because bonds
involving the solute hydrogen atoms were allowed to change. The
pressure was controlled using the Langevin Piston barostat with a
reference pressure of 1 atm and SHAKE\cite{shake77} was used to
constrain bonds involving hydrogen atoms in water. Long range
interactions were treated using particle Ewald mesh PME with a cutoff
of 16 \AA\/ and the Lennard-Jones interactions were switched between
10 \AA\/ and 12 \AA\/. For analysis, every tenth snapshot was saved.\\

\subsection{Intermolecular Interactions}
The present work uses and evaluates different representations of the
bonded and electrostatic interactions. For one, the
CGenFF\cite{cgenff:2020,cgenff:2024} parametrization, consistent with
the TIP3P water model was used for all solutes. Secondly, the ESP of
the solutes was used to obtain off-centered, minimal distributed
charges (MDCM).\cite{MM.mdcm:2017} Thirdly, the total energy function
of the solutes was represented as a machine-learned PES using the
PhysNet\cite{MM.physnet:2019} neural network architecture. Such a
representation replaces all terms from an empirical energy function
except for the van der Waals terms for which initially those from the
CGenFF parametrization were used. Finally, the fourth energy function
combined the PhysNet representation whereby the atom-centered
fluctuating charges from PhysNet were replaced by the off-centered
MDCM charges.\\

\noindent
To improve the representation of the electrostatic potential (ESP)
around a molecule beyond that provided by atom-centered point charges,
models using off-centered charges or multipole moments have been
devised.\cite{Stone:1981,MM.mtp:2012,MM.mtp:2013} Alternatively,
point-charge based models such as the distributed charge model (DCM)
or minimal DCMs (MDCM) were developed and
tested.\cite{MM.dcm:2014,MM.mdcm:2017} Such models are based on the
notion that every higher-order multipole can be replaced by closely
spaced point charges around an interaction site. For example, an
atomic dipole moment corresponds to a closely spaced pair of one
positive and one negative charge.\\

\noindent
The reference ESP from electronic structure calculations was obtained
on a grid of $M$ reference points $(x_m,y_m,z_m), m = 1 \dots M$ from
the electron density of the converged wavefunction using the cubegen
utility.\cite{gaussian16} This reference ESP was represented as a
minimal DCM model (MDCM) using $N$ off-center charges fitted to
reproduce the reference ESP with root mean squared error ${\rm
  RMSE}_N$ using differential
evolution.\cite{Storn.DE:1997,MM.dcm:2014,MM.mdcm:2017,MM.mdcm:2020}
Initially, models with $N_1 \leq N \leq N_2$ MDCM charges were
generated. The lowest RMSE$_N$ for each value of $N$ is recorded. If
RMSE$_N$ is constant for the largest values of $N =
[N_2, \dots , N_2-3]$ the model with the lowest RMSE$_N$ is
used. Otherwise, $N$ was further increased and DE was repeated until
the RMSE$_N$ converged to within 1 kcal/mol.\\

\subsection{Intramolecular Interactions}
For the bonded terms, a PhysNet model was trained to best describe the
total energy of each molecule as a function of
geometry.\cite{MM.physnet:2019} PhysNet is a message passing
NN\cite{gilmer:2017} that falls under the family of graph neural
network.\cite{scarselli:2008} Using energies, forces and dipole
moments for $N$ training structures, the loss function
\begin{equation}
  \label{loss}
\begin{aligned}
\mathcal{L} &= w_E \left| E - E^{\text{ref}} \right| + \frac{w_F}{3N}
\sum_{i=1}^{N} \sum_{\alpha=1}^{3} \left| -\frac{\partial E}{\partial
  r_{i,\alpha}} - F^{\text{ref}}_{i,\alpha} \right| \\ &+ w_Q \left|
\sum_{i=1}^{N} q_i - Q^{\text{ref}} \right| + \frac{w_p}{3}
\sum_{\alpha=1}^{3} \left| \sum_{i=1}^{N} q_i r_{i,\alpha} -
p^{\text{ref}}_{\alpha} \right| + \mathcal{L}_{\text{nh}}.
\end{aligned}
\end{equation}\\
was minimized using the Adam optimizer.\cite{kingma:2014,reddi:2019}
Here, $E^{\text{ref}}$, $F^{\text{ref}}_{i,\alpha}$, $Q^{\text{ref}}$,
and $p^{\text{ref}}_{\alpha}$ are the reference energies, the
Cartesian components of the force on each atom $i$, the total charge,
and the Cartesian components of the reference dipole moment
respectively. The
hyperparameters\cite{MM.physnet:2019,MM.physnet:2023} $w_i$ $i \in \{
E, F, Q, p \}$ differentially weigh the contributions of the loss
function and were $w_E = 1$, $w_F \sim 52.92$, $w_Q \sim 14.39$ and
$w_p \sim 27.21$, respectively. Finally, the term
$\mathcal{L}_{\text{nh}}$ is a ``nonhierarchical penalty'' that serves
as a regularization to the loss function.\cite{MM.physnet:2019} In the
following the PhysNet representations were used in two ways. The first
employed the full model that comprises representations of the bonded
terms and fluctuating point charges whereas for the second usage the
fluctuating point charges were replaced by the MDCM models.\\

\noindent
Reference data for the initial training were generated from MD
simulations using semi-empirical tight binding
GFN2-xTB\cite{bannwarth.XTB:2019} at temperatures $T \in {100, 300,
  1000, 1500, 2000}$ K using ASE\cite{larsen:2017}. From simulations
at each temperature, $\sim 5000$ structures were extracted. Therefore
a total of 25000 geometries was obtained. For each geometry a single
point calculation was carried out at the MP2/6-31G(d,p) level of
theory using MOLPRO2020.\cite{werner.molpro:2020}\\

\noindent
The datasets for the X-Benzenes (X = H, F, Cl, Br) were generated
following a comparable procedure. However, two additional sets of
geometries were generated (at 200 K and 2500 K). Therefore, a total of
35000 structures was generated from MD simulations using the
semiempirical GFN2-xTB model.\cite{bannwarth.XTB:2019} For each
geometry a single point calculation was carried out at the
MP2/6-31G(d,p) level of theory using
MOLPRO2020\cite{werner.molpro:2020}. Therefore, for all the X-Benzenes
(X = H, F, Cl, Br) the total data set contained $\sim 35000$
structures, energies and forces.\\

\noindent
Reference ESPs for all compounds considered were calculated using
Gaussian16 and the cubegen facility.\cite{gaussian16} The data for the
X-Benzenes (where X= H, F, Cl, Br) was calculated at the
PBE1PBE/aug-cc-pvdz level of theory and for the three Cl-PhOH they
were calculated at the MP2/aug-cc-pVTZ level of theory because
PBE1PBE/aug-cc-pvdz calculations did not give any discernible sigma
hole.\\

\subsection{Analysis and Observables}
The radial distribution function $g(r)$ and infrared spectra
$I(\omega)$ were obtained from MD simulations of the hydrated
compounds. All distribution functions $g(r)$ were calculated using
MDAnalysis.\cite{Agrawal.MD_An:2011,beckstein.MD_An:2016} To obtain
the infrared spectrum $I(\omega)$, the dipole moment time series
$\vec{\mu}(t)$ of the solute was computed from the simulations. For MD
using the CGenFF force field, $\vec{\mu}(t)$ was also obtained from
MDAnalysis\cite{Agrawal.MD_An:2011,beckstein.MD_An:2016} whereas for
the other three energy functions (MDCM, PhysNet and PhysNet+MDCM),
$\vec{\mu}(t)$ was obtained from CHARMM and pyCHARMM, respectively.\\

\noindent
From the dipole moment time series $\vec{\mu}(t)$, the autocorrelation
function, $C_{\mu}(t) = \langle vec{\mu}(0) \vec{\mu}(t) \rangle$ was
calculated. Fourier transformation of $C_{\mu}(t)$ using a Blackman
window yields the infrared spectrum
\begin{equation}
\label{IR2}
    I(\omega) = \int e^{-i \omega t} C_{\mu}(t) dt
\end{equation}
which was multiplied by a quantum correction factor $(1 - e^{-\beta
  \hbar \omega})$ to obtain the final signal.\cite{kumar:2004} Here,
$\beta = \frac{1}{k_B T}$, $k_B$ is the Boltzmann constant, $T$ is the
temperature in K, and $\hbar$ is the reduced Planck constant.\\

\noindent
Hydration free energies $\Delta G_{\mathrm{hyd}}$ for all compounds in
water were computed from thermodynamic
integration.\cite{straatsma:1988,MM.cn:2013} For this, the solute was
sampled in the gas phase and in pure water. The condensed-phase
simulations were carried out in the $NpT$ ensemble with 1683 water
molecules, see above. Simulations with 24 different coupling
parameters $\lambda \in (0, 1)$ for electrostatic and vdW interactions
were carried out, respectively. Initial conditions for these
simulation were taken from an unbiased simulation, equilibrated for
50\,ps with the respective coupling parameter $\lambda$ and run for
another 200\,ps for statistical sampling. The total hydration free
energy was then accumulated from
\begin{equation}
     \Delta G_{\mathrm{hyd}} = \sum_\lambda [ (
       H_\mathrm{solv}^\mathrm{elec}(\lambda) -
       H_\mathrm{gas}^\mathrm{elec}(\lambda) ) + (
       H_\mathrm{solv}^\mathrm{vdW}(\lambda) -
       H_\mathrm{gas}^\mathrm{vdW}(\lambda) ) ] \Delta \lambda
\label{eq:hyd}
\end{equation}

\section{Results}

\subsection{Validation of the PhysNet Models}
\begin{figure}[h!]
\centering
\includegraphics[width=1.0\textwidth]{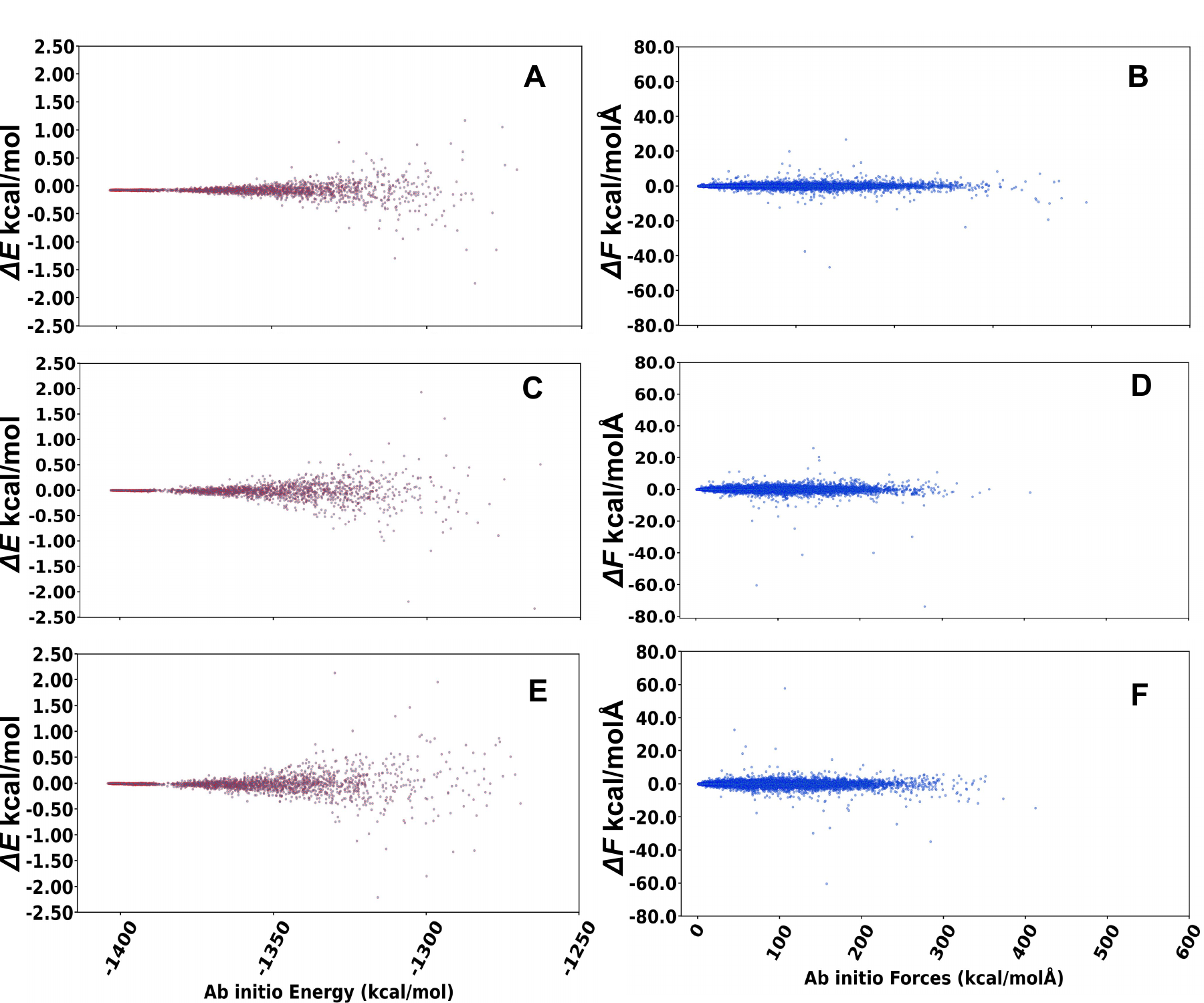}
\caption{Comparison of PhysNet prediction for the test set (2500
  structures) against reference MP2/6-31G(d,p) energies and forces for
  $p-$Cl-PhOH (panels A, B), $m-$Cl-PhOH (C, D), and $o-$Cl-PhOH (E,
  F). Left and right hand columns are for energies and the magnitude
  of the forces, respectively.}
\label{fig1:phenol-EF} 
\end{figure}

The trained PhysNet models were validated with respect to energy,
forces and Normal mode analysis. The performance for the Cl-PhOHs is
reported in Figure \ref{fig1:phenol-EF}. Among the 3 molecules,
$p$-chlorophenol had the highest mean absolute MAE$(E)$ error with
0.10 kcal/mol and 0.07 kcal/mol in test ($\sim 5000$ structures) and
training sets, respectively. The largest RMSE$(E)$ also occurred for
$p$-chlorophenol which were 0.37 kcal/mol and 0.08 kcal/mol in test
and training sets. Compared to this, the maximum absolute error
mAE$(E)$ in the test and training sets were 4.86 kcal/mol and 0.38
kcal/mol for $o-$Cl-PhOH and $m-$Cl-PhOH, respectively. The reported
energetics spanned over a range of 250 kcal/mol and performed within
acceptable range of errors except for a few outliers at higher
energies which correspond to more pronounced molecular distortions. In
terms of forces the highest MAE$(F)$ in the test and training sets
were recorded for $o-$Cl-PhOH with 0.39 kcal/mol/\AA\/ and 0.02
kcal/mol/\AA\/, respectively. This is consistent for RMSE as well
with, $\sim 0.03$ kcal/mol/\AA\/ and 1.05 kcal/mol/\AA\/ for the test
and training sets. The reported ranges of errors for the Halobenzene
are comparable and consistent with earlier work on similar
compounds\cite{MM.physnet:2023} which further validates the present
ML-PESs.\\

\begin{figure}[t]
\centering
\includegraphics[width=1.0\textwidth]{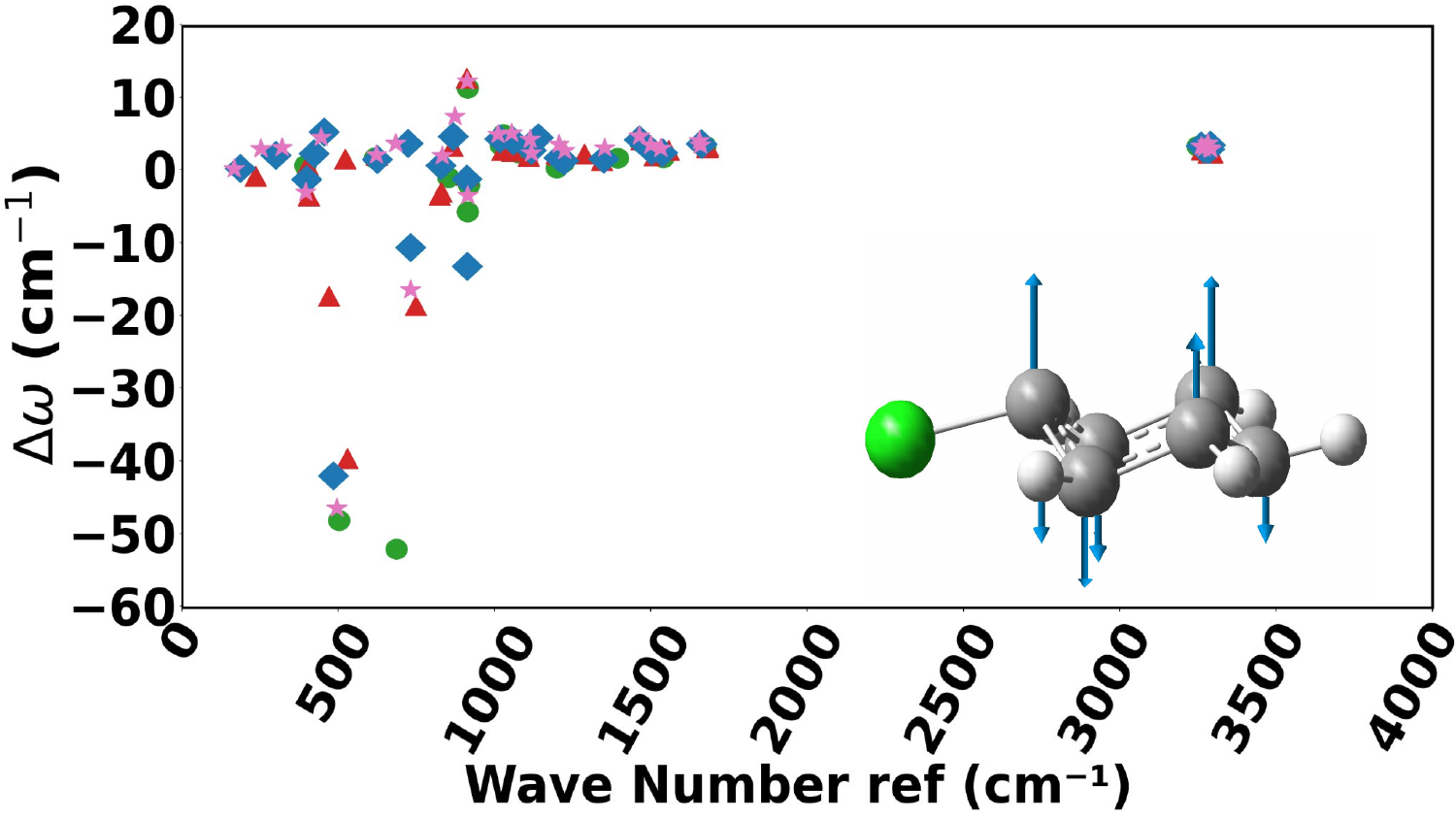}
\caption{Normal mode analysis for X-Bz (where X = H, F, Cl,
  Br). Depending on the frequency $\omega$ of the normal mode the
  difference between {\it ab initio} and harmonic frequencies from
  PhysNet, $\Delta \omega$, differs. The molecular structure shown is
  Cl-Bz and the normal mode vector is that with the maximum deviation
  $\Delta \omega$. Circles, triangles, diamonds, and stars are for
  H-Bz, F-Bz, Cl-Bz, and Br-Bz, respectively.}
\label{fig2:NM-halo}
\end{figure}

\noindent
Another validation of the PESs is afforded by considering the normal
mode frequencies. The average difference between all reference {\it ab
  initio} and predicted normal mode frequencies for Cl-PhOH was
$\Delta \omega \sim \pm 5$ cm$^{-1}$ with a maximum error of $\Delta
\omega \sim 20$ cm$^{-1}$, see Figure \ref{sifig:nmphenol}. This is a
significant improvement over an earlier study\cite{MM.physnet:2023}
for which the maximum $\Delta \omega$ was $ \sim \pm 100$
cm$^{-1}$. The improvement can be attributed to including a larger
number of near-equilibrium structures from MD sampling at 200 K, 300
K, and 1000 K. On the other hand, for the X-Bz, see Figure
\ref{fig2:NM-halo}, the maximum error is $\Delta \omega \sim 50$
cm$^{-1 }$ whereas on average, again, $\Delta \omega \sim \pm 5$
cm$^{-1}$. For both species, see Figures \ref{sifig:nmphenol} and
\ref{fig2:NM-halo}, an increase in $\Delta \omega$ for the lower
frequency ranges was observed. The motion for which the difference
between reference and model harmonic frequency is largest is the ring
out-of-plane bending mode, see Figure \ref{fig2:NM-halo} which was
also the outlier in previous studies. However, despite including
additional samples along this normal mode in the training the
deviations persisted.\\

\subsection{Quality of the MDCM Electrostatic Models}
\begin{figure}[htbp!] 
\centering
\includegraphics[width=1.0\textwidth]{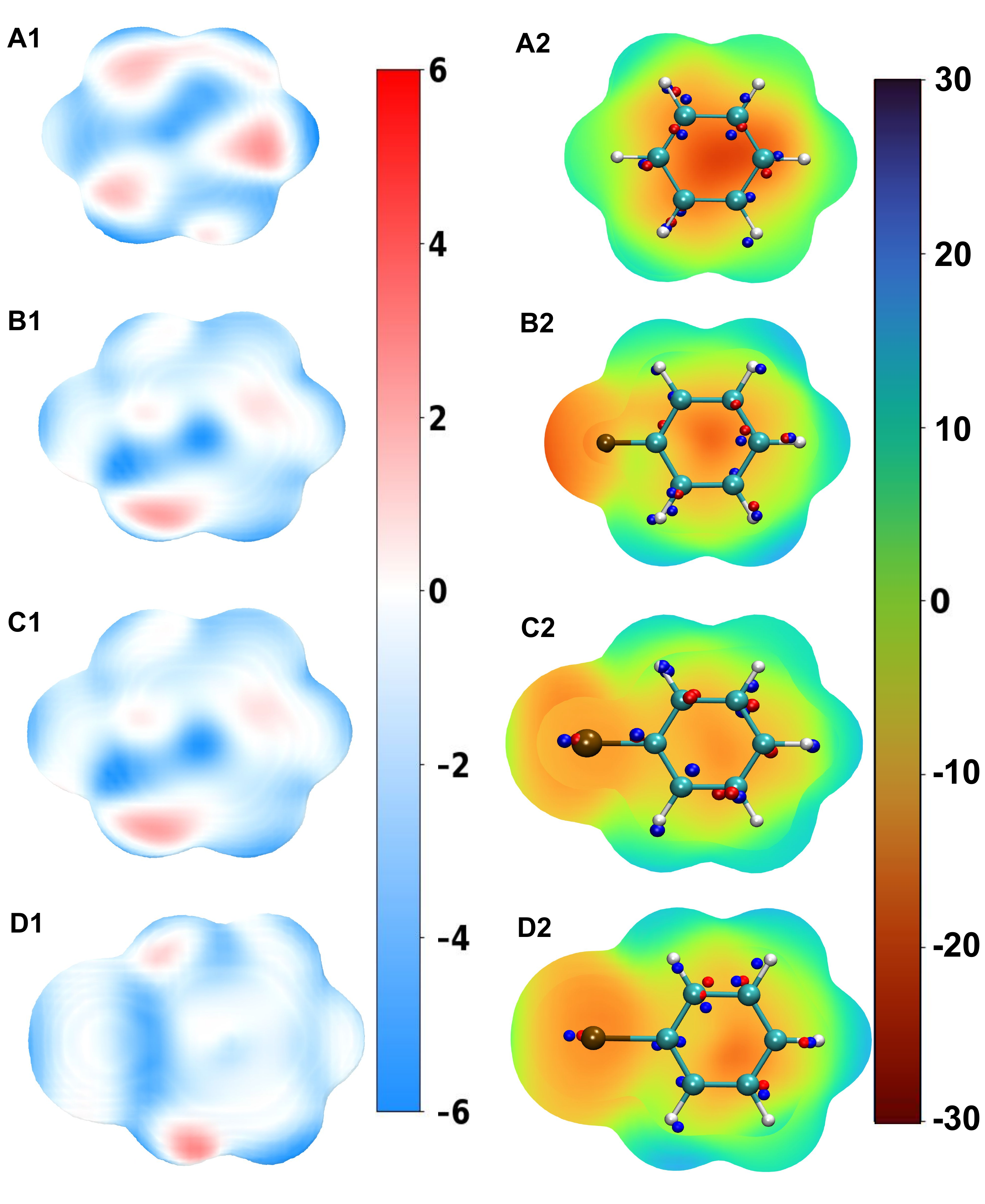}
\caption{Difference ESP ($\Delta {\rm ESP}$) between the reference
  (PBE1PBE/aug-cc-pvdz) and MDCM-generated ESP (left column) and the
  ESP from the MDCM models (right column) for X-Bz (X = H, F, Cl, Br
  from top to bottom) on the 0.001 isosurface. In panels A2 to D2 the
  positions of the positive and negative MDCM charges are reported as
  small blue and red spheres. The geometries are the respective
  equilibrium structures at the PBE1PBE/aug-cc-pvdz level. The average
  errors for the close, mid, and long ranges are reported in Table
  \ref{tab:errors}.}
\label{fig:esp1} 
\end{figure}

\begin{figure}[htbp!] 
\centering
\includegraphics[width=1.0\textwidth]{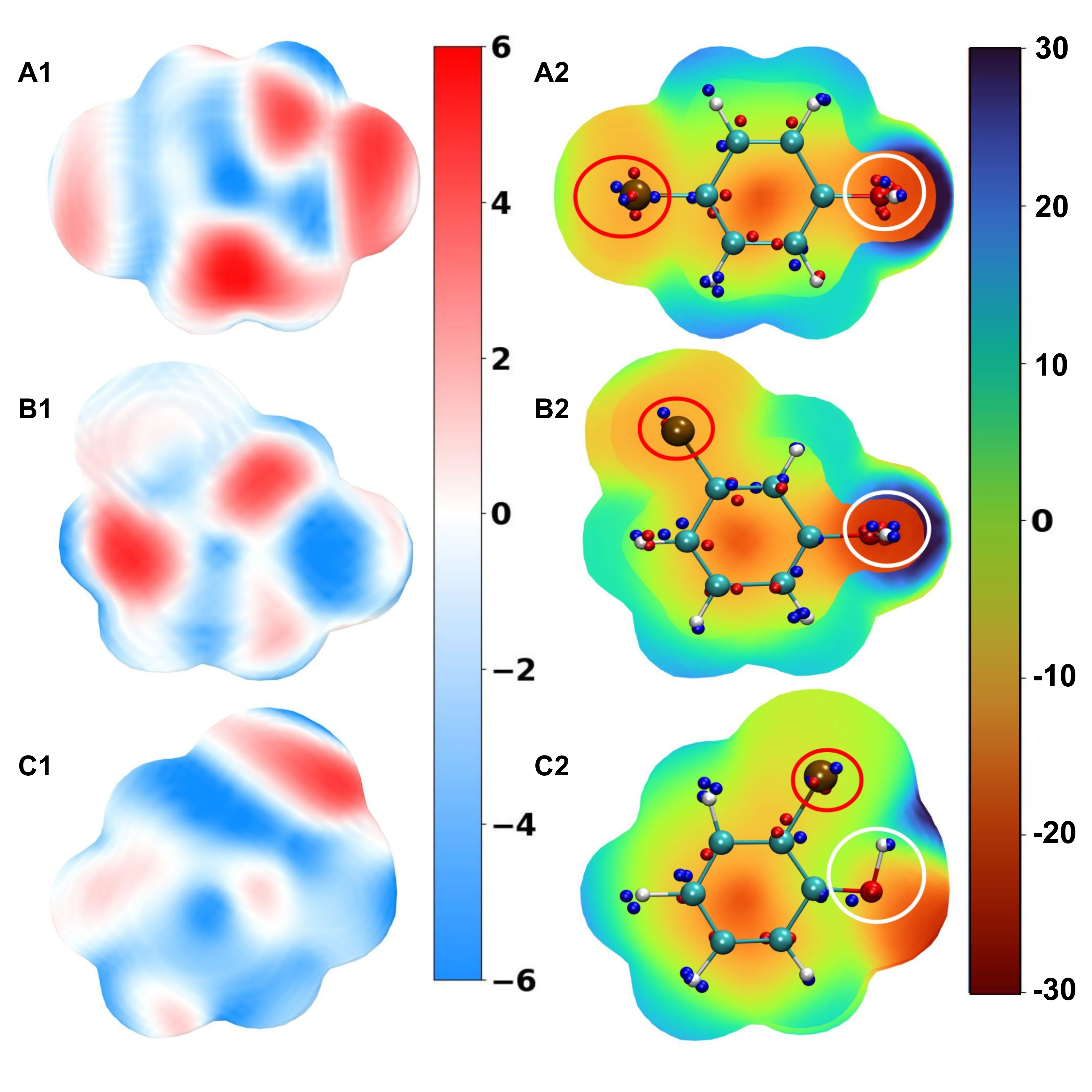}
\caption{Left column: Difference ESP ($\Delta {\rm ESP}$) between the
  {\it ab initio} reference (MP2/aug-cc-pVTZ) and MDCM-generated
  ESP. Right column: the total ESP from the MDCM models. Results for
  $p-$, $m-$, and $o-$Cl-PhOH (from top to bottom) are reported on the
  0.001 isosurface. In panels A2 to C2 the positions of the positive
  and negative MDCM charges are reported as small blue and red
  spheres. The structures used in this Figure corresponds to that for
  which the RMSD compared with the reference ESP is lowest. The
  average errors for the close, mid, and long ranges are reported in
  Table \ref{tab:errors}.}
\label{fig:esp2} 
\end{figure}

The differences between the {\it ab initio} reference ESP and that
from the MDCM models are reported in the left hand columns in Figures
\ref{fig:esp1} and \ref{fig:esp2} for the four X-Bz and the three
Cl-PhOH, respectively. The right hand columns report the positions of
the conformationally averaged charge positions together with the ESP
generated by the MDCM models. In Figure \ref{fig:esp1} the
$\sigma-$hole (blue density along the C-X axis) on the X-site (X = H,
F, Cl, Br) grows as the size of the halogen atom increases. It is also
found that the negative density around the C-X bond grows in magnitude
in going from F to Br. The average total errors range from 0.70 to
0.48 kcal/mol in going from H-Bz to Br-Bz using 19 MDCM charges. This
is consistent with earlier work that reported an error 0.35 kcal/mol
for Br-Bz for which, however, the density was determined at the
B97D3/aug-cc-pVTZ level of theory.\cite{MM.mdcm:2017} Also, the
decreasing error in going from the close to the long range agrees with
earlier work. Overall, the performance of the 19-charge MDCM models
for X-Bz is satisfactory.\\

\begin{table}[h]
    \centering
    \caption{Quality of the MDCM models for X-Bz and $p-$, $m-$,
      $o-$Cl-PhOH. RMSE and maximum errors (in brackets) in
      (kcal/mol)/$e$ for each model are reported. $N_{\rm MDCM}$ is
      the total number of MDCM charges. Close, mid, and long range
      correspond to the ESP in regions within $1.6 \sigma$, between
      $1.6\sigma$ and $2.2\sigma$, and beyond $2.2\sigma$ away from
      the nuclei where $\sigma$ are the van der Waals radii of the
      respective atoms.\cite{Bondi:vanderwaals-1964}}
    \renewcommand{\arraystretch}{1.5}
    \begin{tabular}{|l|c|>{\centering\arraybackslash}m{2.2cm}|>{\centering\arraybackslash}m{2.2cm}|>{\centering\arraybackslash}m{2.2cm}|>{\centering\arraybackslash}m{2.2cm}|}
        \hline
        \multirow{2}{*}{Molecule} & \multirow{2}{*}{$N_{\rm MDCM}$} & \multicolumn{4}{c|}{Errors (ESP$_{\text{ref}}$ - ESP$_{\text{MDCM}}$) kcal/mol} \\ \cline{3-6}
         &                      & Close range      & Mid range       & Long range       & Total      \\ \hline
        H-Bz                   & 19                   & 1.43 (7.37)      & 0.44 (2.18)     & 0.15 (0.83)      & 0.70 (7.37)      \\ \hline
        F-Bz             & 19                   & 1.26 (10.22)     & 0.32 (1.44)     & 0.11 (0.59)      & 0.62 (10.23)      \\ \hline
        Cl-Bz             & 19                   & 1.24 (7.18)      & 0.34 (2.09)     & 0.12 (0.92)      & 0.52 (7.18)      \\ \hline
        Br-Bz              & 19                   & 1.35 (7.10)      & 0.36 (2.34)     & 0.12 (1.02)      & 0.48 (7.10)      \\ \hline
        $p-$Cl-PhOH   & 35                   & 2.37 (13.49)     & 1.79 (4.52)     & 1.36 (2.62)      & 1.67 (13.49)      \\ \hline
        $m-$Cl-PhOH   & 36                   & 1.80 (8.92)      & 0.79 (3.29)     & 0.41 (1.56)      & 0.86 (8.92)      \\ \hline
        $o-$Cl-PhOH   & 34                   & 2.13 (12.9)      & 0.77 (3.21)     & 0.34 (1.16)      & 0.96 (12.90)      \\ \hline
    \end{tabular}
    \label{tab:errors}
\end{table}

\noindent
For the three Cl-PhOH substitutions ($o-$, $m-$, and $p-$Cl-PhOH) the
reference and MDCM-based ESPs are compared in Figure \ref{fig:esp2}
(left hand column) for one of the structures used for obtaining the
conformationally averaged MDCM models. The right-hand column clearly
display the $\sigma-$holes from the MDCM models for $p-$ and
$m-$Cl-PhOH as regions of positive charge density along the C-Cl axis
whereas for $o-$Cl-PhOH the $\sigma-$hole is not particularly
pronounced. This may be caused by the proximity of the -OH group. It
is interesting to note that the centered negative charge distribution
around the -OH group for $p-$ and $m-$Cl-PhOH is shifted away from the
halogenated site for $o-$Cl-PhOH. This leads to a $\sigma-$hole for
$o-$Cl-PhOH that is moved away from the C-Cl axis. The total errors
for different regions away from the molecule (close, mid, long range
and total) are summarized in Table \ref{tab:errors}. The overall
performance of the MDCM models is around 1 kcal/mol on average except
for $p-$Cl-PhOH for which the error is 1.6 kcal/mol. Compared with
X-Bz the average errors are about a factor of two larger for the three
Cl-PhOH. Further reduction of the errors may be achieved by using a
flexible or kernelized MDCM model.\cite{MM.fmdcm:2022,MM.kmdcm:2024}\\

\subsection{Structural Dynamics in Solution}
Using the PhysNet and MDCM models the structural dynamics in solution
was investigated next. Simulations were carried out using the 4
different energy functions: conventional CGenFF (using point charges),
CGenFF but point charges replaced by the MDCM models, PhysNet, and
PhysNet+MDCM where the fluctuating PhysNet charges are replaced by
MDCM charges. All these simulations were carried out using
conventional CHARMM (CGenFF, MDCM) or pyCHARMM with provisions for
ML-PESs (PhysNet and
PhysNet+MDCM).\cite{Charmm-Brooks-2009,pycharmm:2023,MM.physnet:2019,MM.physnet:2023,MM.charmm:2024}
The van der Waals parameters in all four cases were those from the
CGenFF force field.\\

\begin{figure}[ht!]
\centering
\includegraphics[width=1.0\textwidth]{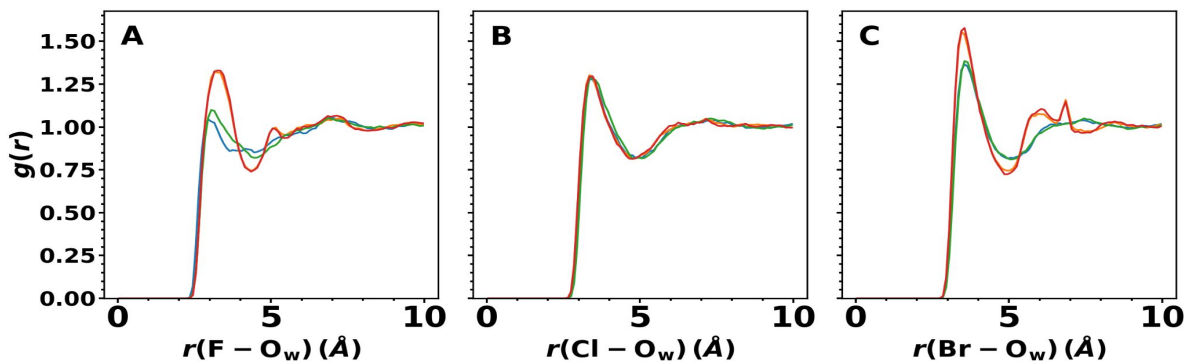}
    \caption{Radial distribution functions $g_{\rm X-O_W}(r)$ between
      the halogen and the surrounding water-oxygen atoms for F-Bz,
      Cl-Bz, and Br-Bz (panels A to C). The blue, orange, green and
      red traces correspond to simulations using the CGenFF, MDCM,
      PhysNet and PhysNet+MDCM energy functions. The orange and red
      traces using the MDCM models are nearly superimposed whereas the
      atom-centered models (CGenFF and PhysNet) differ slightly for
      F-Bz but approach one another for Cl-Bz and Br-Bz.}
\label{fig:gr-halo}
\end{figure}

\noindent
The structural dynamics of the hydrated X-Bz and Cl-PhOH solutes were
characterized in terms of pair distribution functions $g(r)$ between
solute and solvent. For X-Bz (X = F, Cl, Br) the $g_{\rm X-O_W}(r)$
between the halogen and the water-oxygen atoms are shown in Figure
\ref{fig:gr-halo} using the different energy functions (CGenFF, MDCM,
PhysNet and PhysNet+MDCM). For F-Bz (Figure \ref{fig:gr-halo}A) the
occupation in the first solvent shell is $\sim 25$ \% larger from the
simulations using the MDCM model together with either CGenFF bonded
parameters or PhysNet. The CGenFF and PhysNet simulations using
atom-centered point charges (fixed vs. fluctuating) yield comparable
$g(r)$. For Cl-Bz, all four models give essentially identical radial
distribution functions whereas for Br-Bz the observations are similar
to F-Bz. It is noted that for X-O$_{\rm W}$ separations larger than 5
\AA\/ the MDCM models feature some overstructuring which is not
present in the atom-centered point charge models. Such overstructuring
is also known from simulations using atom-centered
multipoles.\cite{MM.eutectic:2022}\\

\noindent
Compared with earlier simulations for X-Bz using MTPs with refitted
van der Waals parameters\cite{MM.mtp:2016} it is found that the
increase in occupation for the first solvent shell for F-Bz is
consistent with using a refined charge model such as MDCM. Also, the
finding that for Cl-Bz the $g_{\rm Cl-O_W}(r)$ is largely independent
of the charge model used is mirrored in the present work and
reassuring. Finally, the increase in water occupation in the first
shell for Br-Bz compared with F-Bz and Cl-Bz has been reported from
simulations using MTPs for Br-Bz.\cite{MM.mtp:2016} In conclusion, for
the X-Bz systems the present findings for the solvent structure is
compatible with earlier simulations using atom-centered multipoles on
the same systems.\\

\begin{figure}[ht!]
  \centering
  \includegraphics[width=1.0\textwidth,height=0.4\textheight]{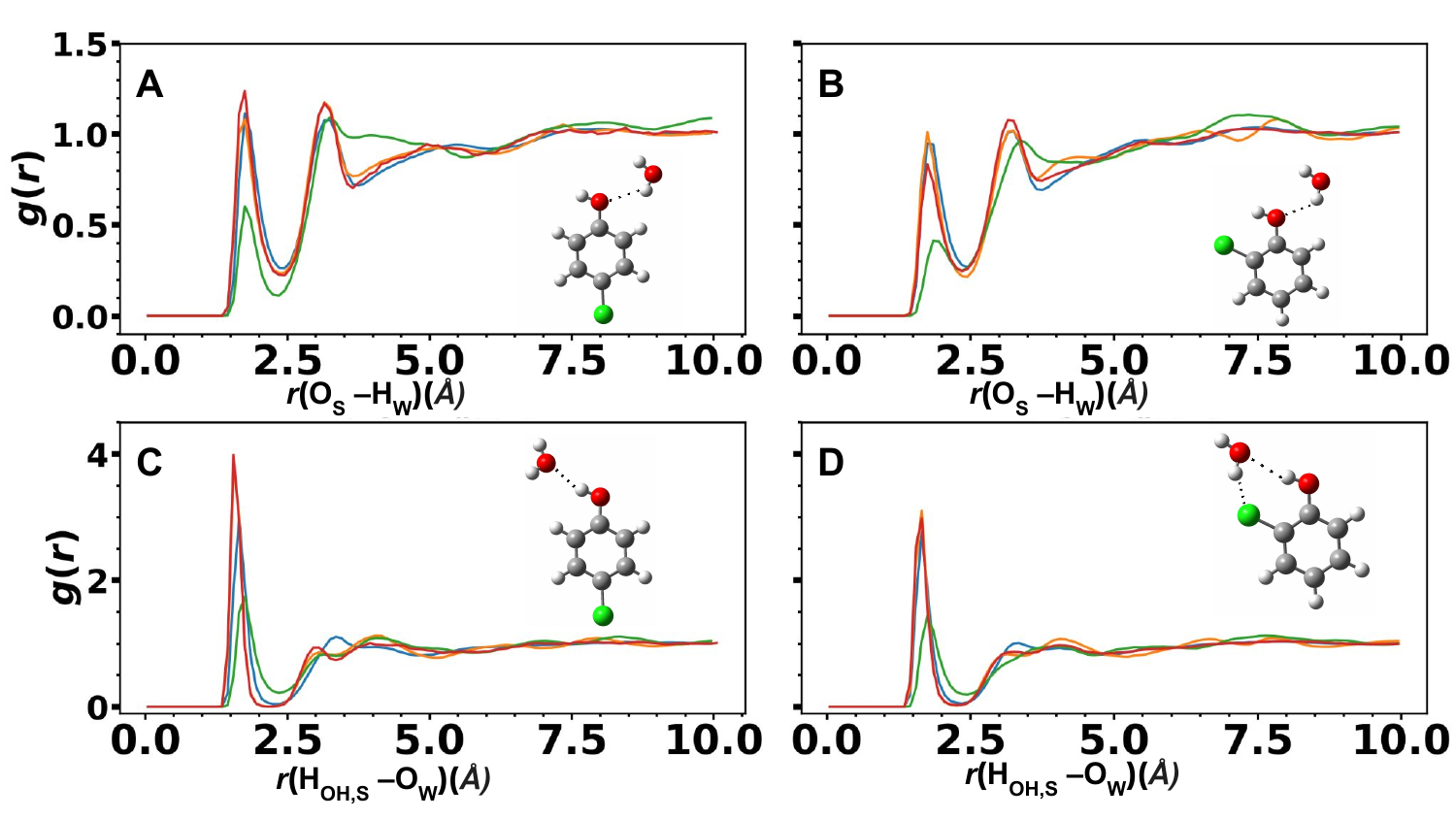}
    \caption{Radial distribution functions between solute (S) and
      water (W) atoms $g_{\rm O_{S}-H_W}(r)$ (top) and $g_{\rm
        H_{OH,S}-O_W}(r)$ (bottom) for $p-$Cl-PhOH (left) and
      $o-$Cl-PhOH (right). The definition of the atom--atom
      separations considered is given in the insets in panels A to
      D. The blue, orange, green and red traces correspond to
      simulations using the CGenFF, MDCM, PhysNet and PhysNet+MDCM
      energy functions. For all $g(r)$ obtained from simulations run
      with PhysNet the intensity of the first peak is considerably
      smaller than for the other three models, see also Figure
      \ref{fig:charge-distribution}.}
\label{fig:gr-phenol}    
\end{figure}

\noindent
For $p-$ and $m-$Cl-PhOH the radial distribution functions $g_{\rm
  O_{S}-H_W}(r)$ and $g_{\rm H_{OH,S}-O_W}(r)$ were considered, see
Figure \ref{fig:gr-phenol}. The water-hydrogen interacting with the
oxygen of the PhOH leads to two prominent peaks at $\sim 1.7$ \AA\/
and $\sim 3.2$ \AA\/ for $p-$ and $m-$Cl-PhOH using the CGenFF, MDCM
and PhysNet+MDCM energy functions, respectively. The positions of the
maxima are nearly identical and the maximum intensities differ by less
than 10 \%. For simulations using the PhysNet model (featuring
fluctuating atom-centered point charges) the first peak is reduced by
$\sim 50$ \% whereas the second peak has comparable intensity but
broadens towards larger separations. A possible rationalization for
this is discussed further below. For the hydrogen bond between the
hydrogen of the PhOH$_{\rm OH}$ and the water oxygen atoms (Figures
\ref{fig:gr-phenol}C and D) simulations using the MDCM models (red and
orange traces) yield nearly identical $g_{\rm O_{PhOH}-O_W}(r)$
whereas those using CGenFF (blue) and PhysNet (green) differ
appreciably. This difference is particularly apparent for $p-$Cl-PhOH
but less evident for $m-$Cl-PhOH.\\

\begin{figure}[htbp!]
  \centering
  \includegraphics[width=1.0\textwidth, height=0.78\textheight]{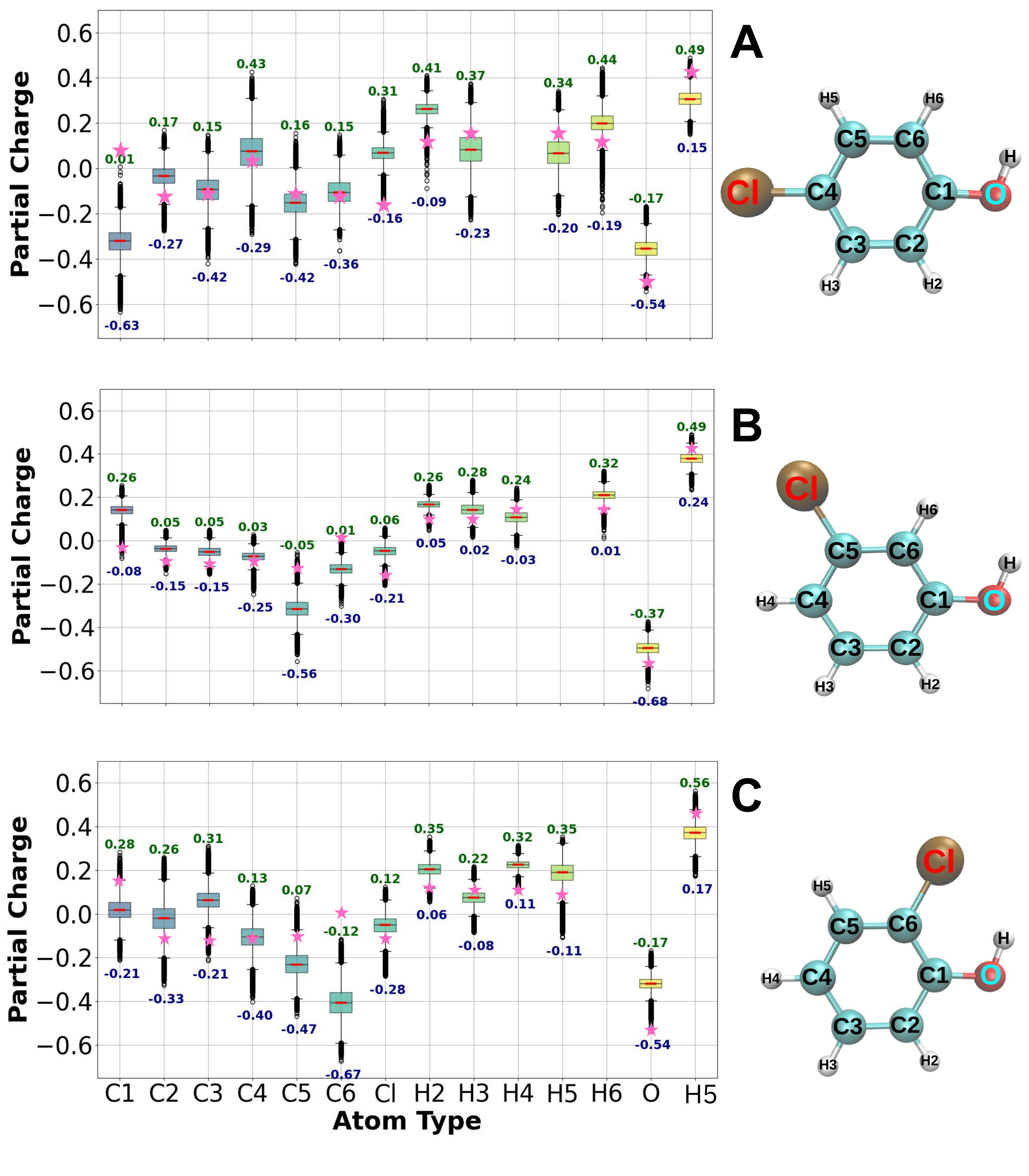}
 \caption{Charge distribution of the conformations for $p-$Cl-PhOH,
   extracted from the pyCHARMM simulation using PhysNet. For a 1 ns
   simulation every 10$^{\rm th}$ step the geometries for 50000
   structures were extracted. The central red line is the median of
   the charge distribution and the colored region represents the 50
   percentile of the charge distribution. The black circles are
   outliers and the lila stars denote the partial charges from
   CGenFF. As the Chlorine atom moves from $p-$ to $o-$ the
   electronegativity of the associated carbon atom (C-Cl) increases
   and the electronegativity on C1 or $\text{C}_{\alpha}$ decreases.}
\label{fig:charge-distribution}
\end{figure}

\noindent
To gain further insights into the reasons why PhysNet leads to the
observed differences for the first hydration shell (Figure
\ref{fig:gr-phenol}), the fluctuating charges from fitting the
molecular dipole moments were analyzed, see Figure
\ref{fig:charge-distribution}. From 1 ns simulations for $p-$, $m-$,
and $o-$Cl-PhOH in solution 50000 structures were extracted in regular
intervals. For these structures the distributions of PhysNet charges
were determined, see Figure \ref{fig:charge-distribution}. For
$p-$Cl-PhOH the range of $q_{\rm Cl}^{\rm PhysNet}$ spans from
$-0.16e$ to $+0.31e$ with a 50 percentile of the median (red bar)
close to 0. This compares with $q_{\rm Cl}^{\rm CGenFF} = -0.15
e$. Hence, the lowest PhysNet charge is consistent with CGenFF but is
larger than the fixed point charge by 0.16$e$. For $m-$Cl-PhOH and
$o-$Cl-PhOH the average $q_{\rm Cl}^{\rm PhysNet}$ shifts
progressively towards the value used in CGenFF whereas the span
covered in PhysNet remains around $\sim 0.3e$ to $\sim 0.4e$. For the
O$_{\rm PhOH}$ the average PhysNet charges for $p-$ and $o-$Cl-PhOH
(Figures \ref{fig:charge-distribution}A and C) are larger by 0.15$e$
and 0.20$e$, respectively, than those in CGenFF. Hence, the average
interaction energy between O$_{\rm PhOH}$ and the surrounding water is
considerably reduced in simulations using PhysNet compared with
CGenFF. This rationalizes the reduced occupation of the first
solvation shell found in Figure \ref{fig:gr-phenol}.\\

\subsection{Infrared Spectroscopy}
\begin{figure}[ht!]
\centering
\includegraphics[width=1.0\textwidth]{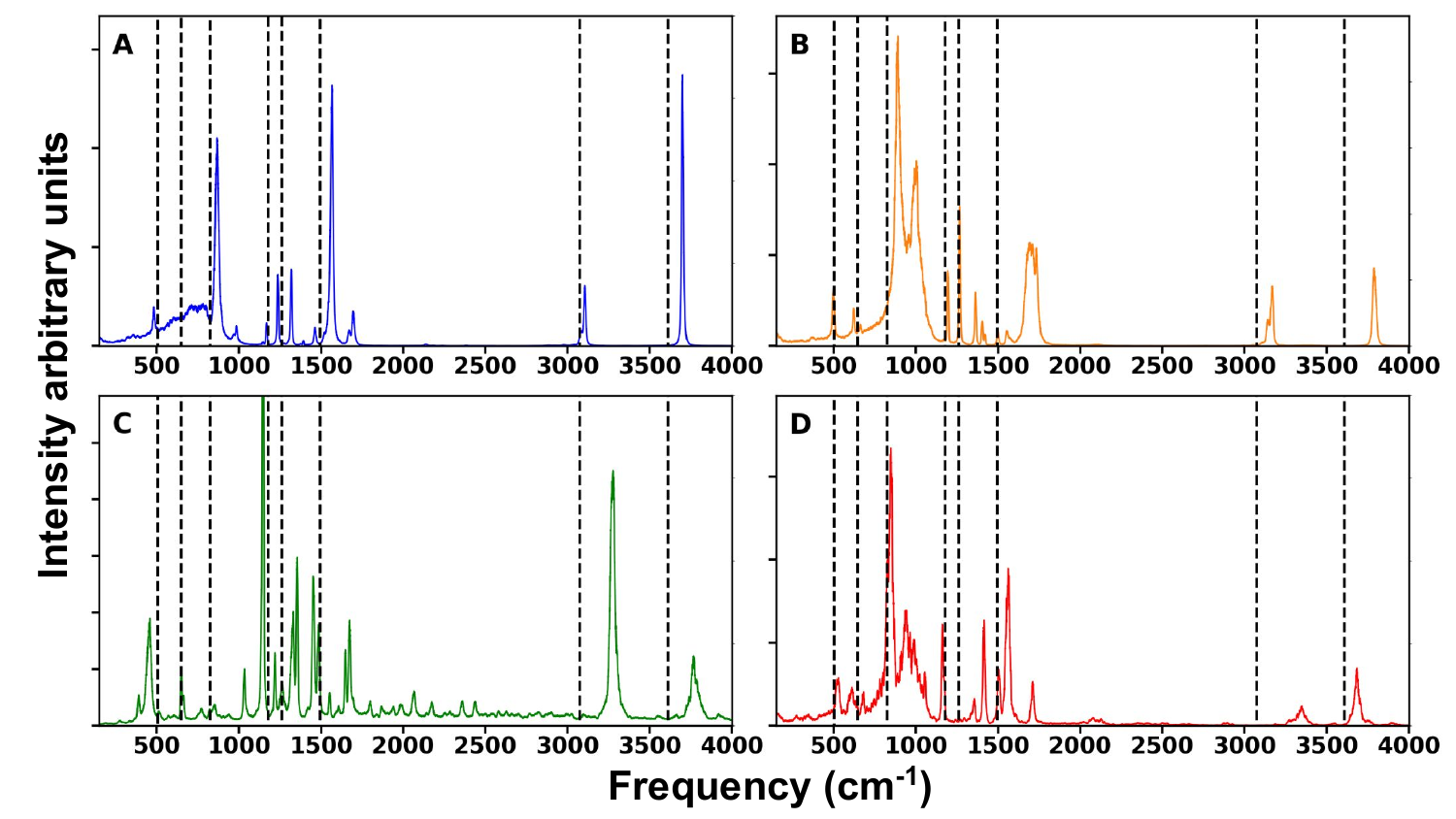}
    \caption{IR spectra for $p-$Cl-PhOH in solvent from MD simulations
      using the CGenFF, MDCM, PhysNet and PhysNet+MDCM energy
      functions, panels A to D. The dashed lines indicate the most
      prominent experimentally measured band maxima which were
      measured in CCl$_4$ and
      cyclohexane.\cite{Zierkiewicz.paraIR:2000} CGenFF almost exactly
      reproduces the CH- and OH-stretch frequencies (because the
      parametrization was chosen accordingly) whereas PhysNet
      overestimates the frequencies due to sampling only the harmonic
      well.\cite{suhm:2020} Also, it is noted that MDCM shifts the CH-
      and OH-bands somewhat to higher frequencies.}
\label{fig:IRphenol}
\end{figure}

Infrared spectroscopy is a valuable experimental tool that provides
molecular-level information. IR spectra for $p-$Cl-PhOH using the four
interaction models are reported in Figure \ref{fig:IRphenol}. The CH-
and OH-stretch-involving vibrations from simulations using CGenFF
(panel A) are in good agreement with experiment. This is primarily
because the parametrization had been chosen accordingly which is an
advantage of empirical energy functions. Using MDCM as the
electrostatic model (panel B) shifts the frequencies slightly to the
blue whereas the two PhysNet models (panels C and D) lead to more
pronounced blue shifts for the CH-stretch whereas the OH-stretch mode
is of comparable quality from PhysNet+MDCM (panel D) as CGenFF. The
shifts observed for the MDCM model are consistent with earlier work on
CO in myoglobin which also reported that multipolar charge
distributions shift IR spectra depending on the strengths of the
interactions.\cite{MM.mbco:2003} For the two PhysNet models the
differences between simulations and experiments are due to i) the
level of theory at which the models were trained and ii) sampling only
the bottom of the well in MD simulations which is a well-understood
effect.\cite{suhm:2020} The pattern of framework modes (below and
around 1000 cm$^{-1}$) is realistically captured from CGenFF
simulations whereas MDCM widens all bands (panel B). PhysNet in panel
C yields a comparably good pattern of vibrations as CGenFF and using
PhysNet+MDCM (panel D) broadens the spectral features around 1000
cm$^{-1}$. More specifically, the bands at 500 cm$^{-1}$, 823
cm$^{-1}$, 1176 cm$^{-1}$, 1259 cm$^{-1}$, 1494, 3073 cm$^{-1}$ and
3609 cm$^{-1}$ are well reproduced using CGenFF, whereas the peaks at
644 cm$^{-1}$ and 1494 cm$^{-1}$ are less well captured. Comparable
observations are made for the simulations using PhysNet. A more
detailed analysis and assignments is not attempted here. It is noted
that a dedicated experimental/computational study of the vibrational
spectroscopy of $p-$F-PhOH showed that the vibrations in the region of
the framework modes (1000 to 1200 cm$^{-1}$ are strongly
mixed.\cite{MM.fphoh:2022} Unequivocal assignment of the vibrations
was not possible in that case either.\\

\noindent
For completeness, computed IR spectra for the X-Benzenes (where X = H,
F, Cl, Br) in water are reported in Figure \ref{sifig:ir-halo}
together with available
experiments.\cite{John.Cl-Bz-IR:2000,ramasamy:2014,Toyozo:1977,zhou:2012}
It is noted that data is only available in pure liquids, KBr pellets,
or in a liquid film but not in
water. Experimentally,\cite{bertie:1994} four infrared bands were
reported for liquid benzene and benzene in the gas phase with center
frequencies at $\sim 675$ cm$^{-1}$, $\sim 1036$ cm$^{-1}$, $\sim
1479$ cm$^{-1}$, and $\sim 3070$ cm$^{-1}$. Using CGenFF (panel A1)
for simulations in water these bands are rather well reproduced also
because the FF was fitted to do so. Replacing the atom-centered point
charges by MDCM widens the absorption band which is consistent with
experiments in liquid benzene. With PhysNet (A3) the absorption is
shifted to the blue which is a consequence of the level of quantum
chemical theory at which the reference data was calculated and the
fact that MD simulations sample the bottom of the
well.\cite{suhm:2020} Finally, switching to MDCM charges (A4) again
widens the spectrum. For the CH-stretch vibrations a slight blue shift
is found for simulations using MDCM (orange) and with PhysNet the
shift to the blue is more appreciable. Again, the reasons for these
findings are identical to those for Cl-PhOH above. No attempt is made
for a detailed assignment of the framework modes for halogenated
benzenes.\\

\subsection{Hydration Free Energies}
Finally, hydration free energies from using the CGenFF, MDCM and
PhysNet models were determined. For PhysNet+MDCM additional code
developments will be required. Using MDCM two different approaches
were chosen. In the first the standard CGenFF van der Waals parameters
were used whereas in the second the van der Waals ranges were allowed
to vary to better reproduce the experimentally observed $\Delta G_{\rm
  hyd}$. Adaptation of the van der Waals ranges is known to be
necessary if the underlying electrostatic model is
changed.\cite{MM.cn:2013,MM.oh:2013,MM.eutectic:2024}\\

\noindent
Using the point-charge-based CGenFF parametrization the hydration free
energies for all compounds considered are in reasonably good
agreement with experiment, see Table \ref{tab:dg}, except for
F-Bz. Typical deviations are 10 \% to 25 \% and most of the rankings
are correctly reproduced. This is to be expected as hydration free
energies were part of the parametrization procedure of the energy
function.\cite{cgenff:2020,cgenff:2024}\\

\noindent
Replacing PCs by MDCMs to capture charge anisotropy but retaining the
remaining parameters of the CGenFF energy function overestimates
$\Delta G_{\rm hyd}$ for X-Bz by about a factor of 2 for halogens and
a factor of 6 for benzene, see Table \ref{tab:dg}. On the other hand,
for the three Cl-PhOH species the performance is rather encouraging,
in particular for $m-$Cl-PhOH. For $p-$ and $o-$Cl-PhOH the hydration
free energy is overestimated by about 25 \% or $\sim 2.5$ kcal/mol. As
is well known, replacing the electrostatic model in an empirical
energy function also requires the van der Waals parameters to be
readjusted. Scaling the $\sigma-$parameters (range) is one possibility
which was followed here. For the X-Bz species, increasing the
vdw-ranges by up to 20 \% yields satisfactory results - certainly
within experimental error bars of $\sim 1$ kcal/mol - whereas for the
Cl-PhOH species scaling by 20 \% improves the performance
considerably. The only exception is benzene for which increasing the
vdw-ranges still leads to an overestimation by a factor of 3. It was
already found previously that in addition to modifying $\sigma-$values
it may also be necessary to decrease the well depth $\epsilon$ of the
van der Waals interactions.\cite{MM.mtp:2013} In addition, the MDCM
models for X-Bz were not conformationally averaged and doing so may
further improve the hydration free energies for benzene.\\

\begin{table}[b!]
\centering
\caption{Hydration free energies $\Delta G_{\text{hyd}}$ for the
  different compounds and energy functions considered in the present
  work. Comparison between point charge model (PC), Minimally
  distributed charge model (MDCM), and PhysNet. The performances were
  benchmarked against experimental
  values.\cite{marenich:2020,abraham:1990,rizzo:2006,vyboishchikov:2021}. $\Delta
  G_{\text{hyd}}$ using MDCM were obtained using the same CGenFF
  Lennard-Jones parameters whereas for $\Delta G_{\text{hyd}}^{*}$ the
  Van der Waals radii were increased by 30 \% for H-Bz, by 20 \% for
  $o-$Cl-PhOH and Cl-Bz, and by 10 \% for all other molecules. Error
  bars were determined from averages over 4 independent
  simulations. Typical experimental errors for neutrals are estimated
  at $\sim \pm 1$ kcal/mol.\cite{mobley:2014}}
\renewcommand{\arraystretch}{1.5}
\begin{tabular}{|>{\centering\arraybackslash}m{2.8cm}|>{\centering\arraybackslash}m{2.0cm}|>{\centering\arraybackslash}m{2.0cm}|>{\centering\arraybackslash}m{2.0cm}|>{\centering\arraybackslash}m{3.0cm}|>{\centering\arraybackslash}m{2.0cm}|}
\hline
\textbf{Molecules}         & $\Delta G_{\rm hyd}$ PC & $\Delta G_{\rm hyd}$ MDCM & $\Delta G_{\rm hyd}^{*}$ MDCM & $\Delta G_{\rm hyd}$ PhysNet & $\Delta G_{\rm hyd}$ Exp \\ \hline
H-Bz                    & $-0.91 \pm 0.33$        & $-4.90 \pm 0.80 $              & $-2.69 \pm 0.79$             & $-1.21 \pm 0.23$               & --0.87                \\ \hline
F-Bz                    & $-1.49 \pm 0.45 $        & $-1.60 \pm 0.51$              & $-1.12 \pm 0.61$             & $-1.12 \pm 0.22$              & --0.78                \\ \hline
Cl-Bz                  & $-1.14 \pm 0.39$        & $-2.91 \pm 0.45$             & $-1.75 \pm 0.77$              & $-1.22 \pm 0.22 $                & --1.12                \\ \hline
Br-Bz                  & $-1.58 \pm 0.46$        & $-2.65 \pm 0.38$              & $-1.96 \pm 0.40$             & $-1.30 \pm 0.23$               & --1.46                \\ \hline
$p-$Cl-PhOH          & $-5.01 \pm 0.74$          & $-9.47 \pm 0.81$              & $-7.50 \pm 0.81$              & $-3.91 \pm 0.42$               & --7.03                \\ \hline
$m-$Cl-PhOH         & $-6.37 \pm 0.70 $            & $-6.80 \pm 0.79$             &  $-6.80 \pm 0.79$               & $-3.68 \pm 0.41$               & --6.62               \\ \hline
$o-$Cl-PhOH        & $-5.85 \pm 0.74$            & $-7.20 \pm 0.52$              & $-5.02 \pm 0.49$              & $-3.01 \pm 0.44$                & --4.55             \\ \hline
\end{tabular}
\label{tab:dg}
\end{table}

\noindent
Finally, $\Delta G_{\rm hyd}$ were also determined from simulations
using PhysNet which is based on fluctuating atom-centered point
charges together with van der Waals parameters from CGenFF. Because
PhysNet returns only the total energy (bonded plus electrostatics) of
the solute and the solute-solvent interactions, scaling of the
energies as per Eq. \ref{eq:hyd} needs to be carried out
differently. Two possibilities were considered in the present work. In
the first the total energy of the solute (bonded plus electrostatics)
and the solute--solvent interactions were scaled from 0 to 1 following
Eq. \ref{eq:hyd}, see Table \ref{tab:dg}. Secondly, only the
solute-solvent interactions were scaled from 0 to 1 but the total
energy of the solute was not scaled ($\lambda = 1$ throughout). The
latter approach leads to more favourable binding (more negative
$\Delta G_{\text{hyd}}$) by $\sim 10$ \%. This yields qualitatively
correct rankings of the hydration free energies. For X-Bz the $\Delta
G_{\text{hyd}}$ are somewhat too uniform for H-, F-, and Cl-Bz but
their magnitude is meaningful. Hydration free energies for the three
Cl-PhOH molecules are too weak by $\sim 50$ \%. Again, for
quantitative hydration free energies the van der Waals parameters will
need to be readjusted because the charge model was modified. This is
akin to replacing point charges in CGenFF by MDCM, see above.\\

\noindent
It is of interest to analyze the hydration free energies together with
the radial distribution functions. This was done for $p-$Cl-PhOH, see
Figure \ref{fig:gr-phenol}C. The height of the first peak $g_{\rm
  H_{OH,S}-O_W}(r)$ decreases in magnitude in going from orange/red
(overlapping) to blue to green, i.e. MDCM-based, PC, and PhysNet
models. The corresponding hydration free energies change from --9.5 to
--5.0 to --3.9 kcal/mol. In other words, the hydration free energies
decrease in concert with the maximum probability to find a hydration
water molecule in the first solvation shell. This is also consistent
with the statistical mechanical interpretation of the pair
distribution function as the average strength of the pair
interaction. Findings for $o-$Cl-PhOH follow related trends when
comparing Figure \ref{fig:gr-phenol}D with the data in Table
\ref{tab:dg}.\\

\noindent
Overall, all charge models yield correct rankings with MDCM and
readjusted van der Waals parameters perform best. For $m-$Cl-PhOH no
adjustment was required whereas for $p-$ and $o-$Cl-PhOH the ranges
need to be scaled by 10 \% to obtain agreement with experiment when
accounting for measurement errors. It is notable that for benzene
using MDCM rather large modifications are needed which indicates that
a charge model that includes charge anisotropy is not suitable and
also not necessary. This can be understood by noting that for H-Bz an
atom centered PC model is expected to suffice for modeling the weak
solvent-solute interactions. Introducing a more elaborate MDCM model
may better model the ESP but for correct thermodynamic properties
(here $\Delta G_{\rm hyd}$) the van der Waals terms need to compensate
any exaggeration in representing the ESP alone. Thus, more elaborate
representations of the ESP can lead to models less suitable for
condensed-phase simulations because balancing the different terms is
challenging. This is consistent with earlier findings for
multipole-based approaches.\cite{MM.oh:2013,MM.eutectic:2024}\\

\section{Conclusions}
The present work assesses the possibility and effects of replacing
specific contributions to an empirical energy function with either
more detailed electrostatics (PC to MDCM) or all bonded terms by a
ML-based representation. This was done for halogen-substituted
benzenes and chlorinated phenol in the gas phase and in solution. As
expected, changes in the electrostatics lead to modified hydration
shells as evidenced in pair distribution functions between solutes and
water. This also impacts hydration free energies for which
modifications in the electrostatics require adaptation of the van der
Waals parameters. In particular for the three Cl-PhOH the agreement
between experiment and simulations using an MDCM model is
encouraging. On the other hand, for benzene (H-Bz) the present MDCM
model requires unusually large van der Waals ranges and additional
modification of van der Waals $\epsilon-$values and/or conformational
averaging for the MDCM model may be required for hydration free
energies in better agreement with experiments. Fluctuating
atom-centered charges as available from PhysNet, on the other hand,
yield models that can be easily brought into agreement with experiment
by slightly scaling the van der Waals parameters. The IR spectroscopy
of Cl-PhOH is quite well captured by all models used whereas for the
halogenated benzenes comparison between experiment and simulations is
more difficult due to the different environments in which the
measurements were carried out. \\

\noindent
The available tools to refine empirical energy functions provide a
basis for more targeted improvement for specific applications. This is
of particular relevance for interpretation and eventually prediction
using atomistic simulations. On the other hand, the present work finds
that empirical energy functions serve as a good starting point for
exploring the structure, dynamics and thermodynamics of hydrated
systems. Improving on them requires high levels of electronic
structure calculations, depends on the observables considered, and
necessitates the analysis and improvement of the entire energy
function, and not only parts of it. One point not addressed in the
present work relates to the water model used. For consistency with the
CGenFF parametrization, the TIP3P model was used here. Recent work on
eutectic liquids investigated changes in the parametrization if
another water model is used.\cite{MM.eutectic:2024} However, the
performance of the final models was comparable after fitting to the
same electronic structure reference data.\\

\section*{Data Availability}
The data accompanying this work is available at
\url{https://github.com/MMunibas/halobenzenes}.

\section*{Acknowledgment}
We thank the Swiss National Science Foundation (grants
200020{\_}219779 and 200021{\_}215088) (to MM), and the University of
Basel for supporting this work.

\bibliography{refs}

\clearpage

\renewcommand{\thetable}{S\arabic{table}}
\renewcommand{\thefigure}{S\arabic{figure}}
\renewcommand{\thesection}{S\arabic{section}}
\renewcommand{\d}{\text{d}}
\setcounter{figure}{0}  
\setcounter{section}{0}  
\setcounter{table}{0}

\newpage

\noindent
{\bf SUPPORTING INFORMATION: Machine Learning-Based Enhancements of
  Empirical Energy Functions: Structure, Dynamics and Spectroscopy of
  Modified Benzenes}

\section{PhysNet Models for X-Bz}
\begin{figure}[ht!] 
\centering
\includegraphics[width=1.0\columnwidth,height=0.75\textheight,
  scale=2]{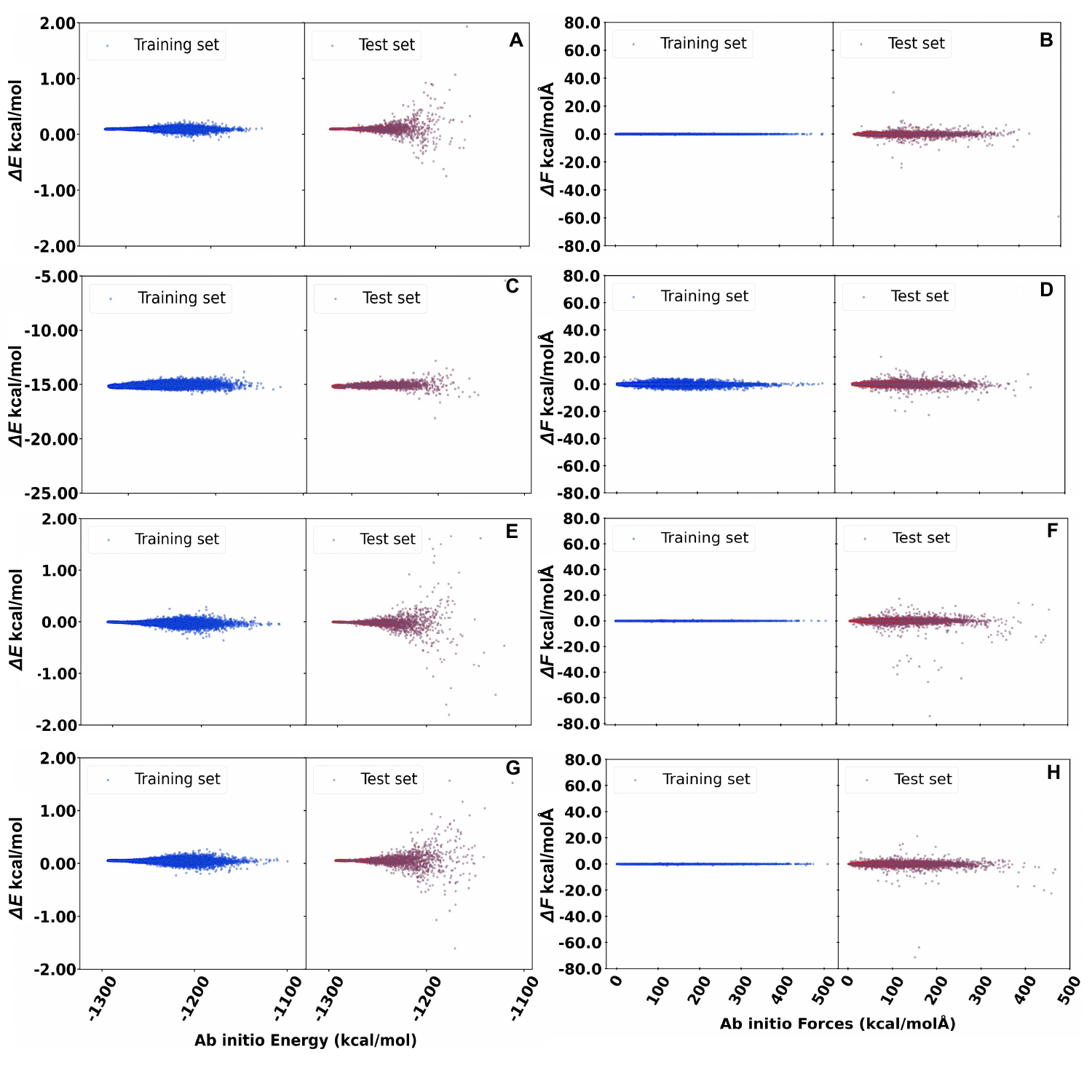}
\caption{Test set error of PhysNet prediction and reference
  MP2/6-31G(d,p) calculations for energies (panels A, C, E, G) and
  forces (B, D, F, H) for X-Bz (X = H, F, Cl, Br) from top to bottom.}
\label{sifig:fig2-HALO-EF}
\end{figure}

\clearpage

\section{Normal Modes for Cl-PhOH}

\begin{figure}[ht!] 
\centering
\includegraphics[width=1.0\columnwidth,height=0.40\textheight,
  scale=2]{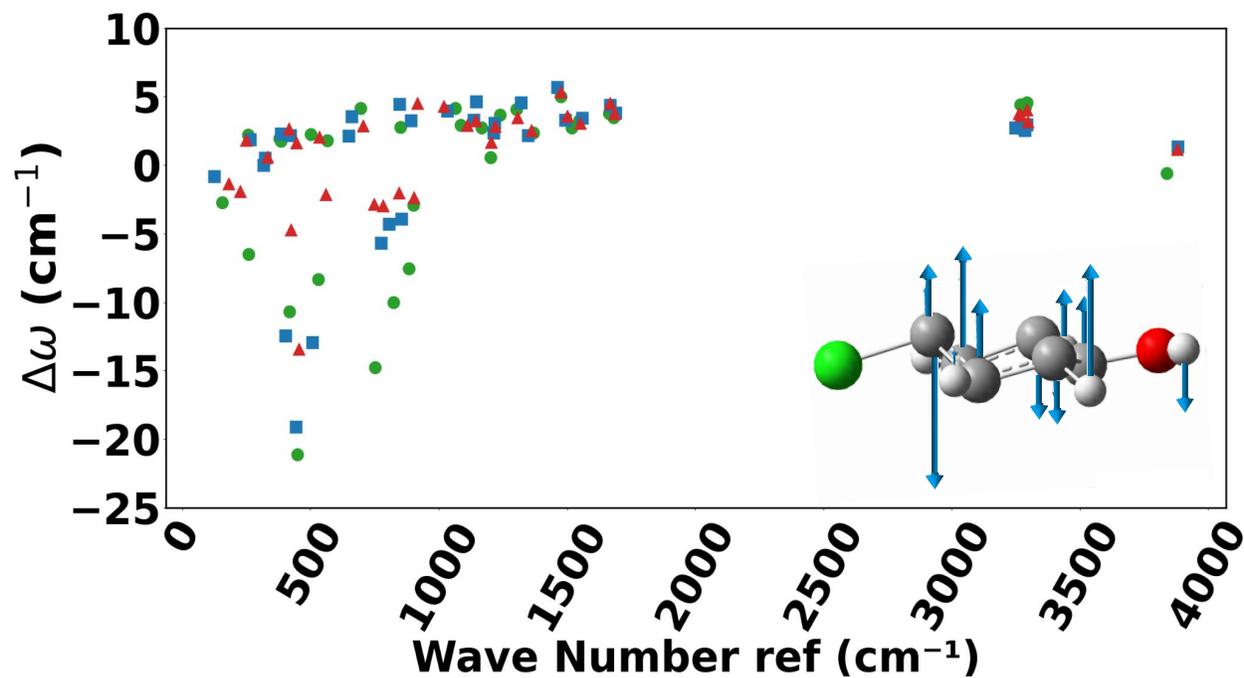}
\caption{Normal mode analysis for $p-$, $m-$, and $o-$ClPhOH (blue,
  red, green symbols). The performance of the trained PhysNet model in
  terms of the difference $\Delta \omega = \omega_{\rm ref.} -
  \omega_{\rm NN}$ between the {\it ab initio} reference and that from
  PhysNet is reported depending on the reference frequency. The inset
  reports the normal mode vibration with the largest error.}
\label{sifig:nmphenol} 
\end{figure}

\clearpage

\section{Infrared Spectroscopy of X-Bz}
\begin{figure}[ht!]
  \centering
  \includegraphics[width=1.0\textwidth]{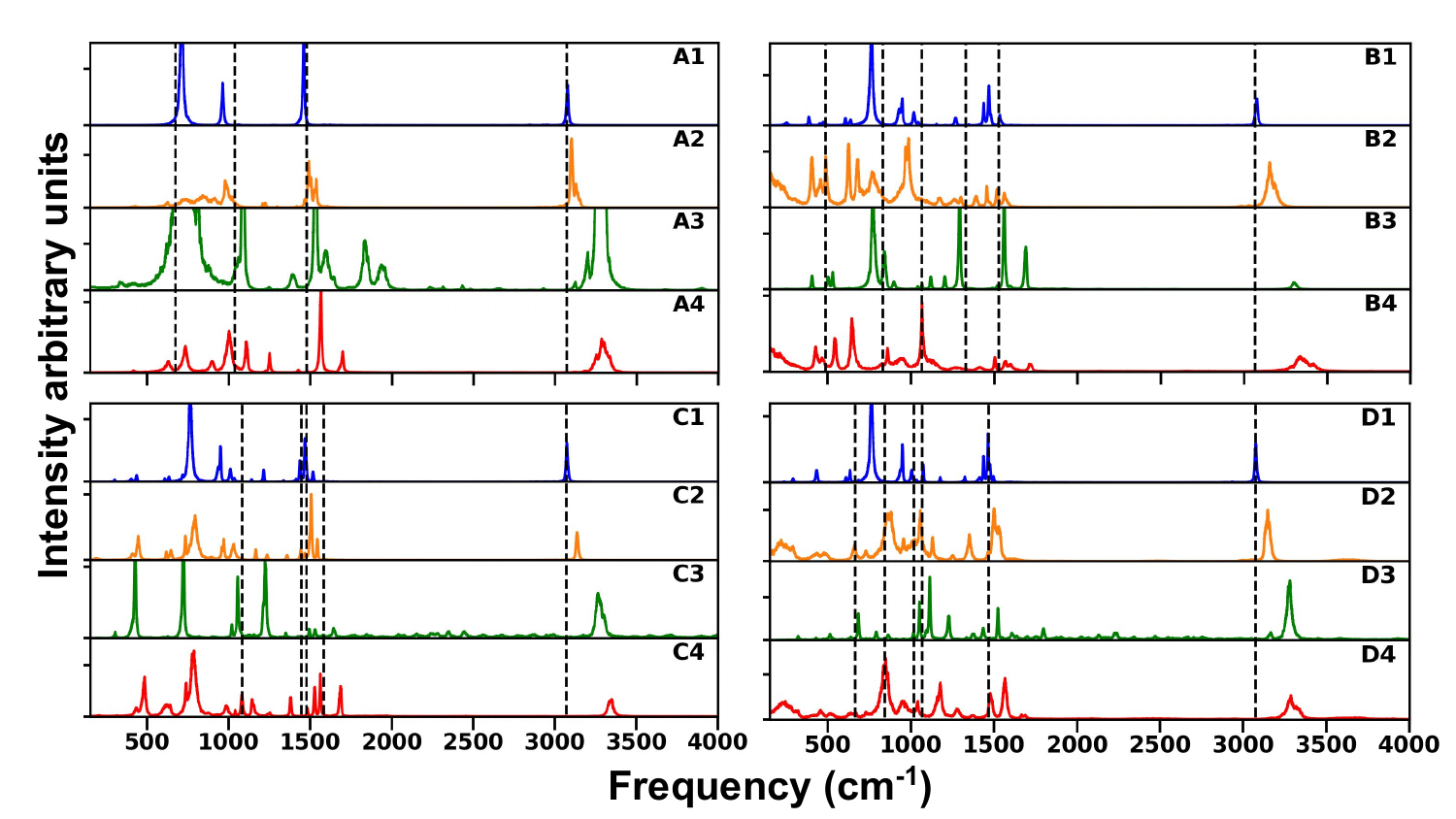}
    \caption{IR spectra for X-Bz in water using the CGenFF, MDCM,
      PhysNet, and PhysNet+MDCM energy functions (blue, orange, green
      red). Panels A to D are for H-Bz, F-Bz, Cl-Bz, and Br-Bz,
      respectively. The black dashed lines are positions of the most
      prominent lines from the experimental IR spectra which were
      recorded in the pure liquid (A), in solid KBr pellets (B), in
      the liquid (C), and in a liquid film
      (D).\cite{John.Cl-Bz-IR:2000,ramasamy:2014,Toyozo:1977,zhou:2012}}
\label{sifig:ir-halo}
\end{figure}

Figure \ref{sifig:ir-halo} reports the calculated IR spectra for the
X-Benzenes (where X = H, F, Cl, Br) in water with available
experiments.\cite{John.Cl-Bz-IR:2000,ramasamy:2014,Toyozo:1977,zhou:2012}
The IR spectrum for H-Bz is shown in Figure \ref{sifig:ir-halo}A using
the CGenFF, MDCM, PhysNet and PhysNet+MDCM energy functions. The
influence of the MDCM charges when using the same bonded parameters
(CGenFF) is clearly visible in panels A1 and A2. The absorption
frequencies remain largely unchanged whereas the intensities clearly
depend on the charge model used. For PhysNet with fluctuating and MDCM
charges similar effects are found: using atom-centered fluctuating
charges (A3) the IR spectrum is comparatively simple whereas replacing
them with MDCM charges the positions of the absorptions remain but the
intensities redistribute. Experimentally,\cite{bertie:1994} four
infrared bands were reported for liquid benzene and benzene in the gas
phase with center frequencies at $\sim 675$ cm$^{-1}$, $\sim 1036$
cm$^{-1}$, $\sim 1479$ cm$^{-1}$, and $\sim 3070$ cm$^{-1}$. Using
CGenFF (panel A1) for simulations in water these bands are rather well
reproduced also because the FF was fitted to do so. Replacing the
atom-centered point charges by MDCM widens the absorption band which
is consistent with experiments in liquid benzene. With PhysNet (A3)
the absorption is shifted to the blue which is a consequence of the
level of quantum chemical theory at which the reference data was
calculated and the fact that MD simulations sample the bottom of the
well.\cite{suhm:2020} Finally, switching to MDCM charges (A4) again
widens the spectrum. For the modes at lower frequency, similar
observations are made.\\

\noindent
For the halogenated benzenes in panels B to D the CH-stretch modes
behave in the same fashion as for H-Bz. Comparisons for the framework
modes between 500 cm$^{-1}$ and 1500 cm$^{-1}$ are more difficult
because no experimental data for the species in water are
available. In general it is noted that the computations yield a larger
number of spectroscopic features than the experiments. At this stage a
rigorous comparison of experimentally measured spectra for F-, Cl-,
and Br-Bz in water is not attempted.\\

\end{document}